\documentclass[11pt]{article}
\usepackage{moriond,epsfig}

\def\Journal#1#2#3#4{{#1} {\bf #2}, #3 (#4)}

\def\PLB{{\em Phys. Lett.}  B}
\def\PRL{\em Phys. Rev. Lett.}
\def\PRD{{\em Phys. Rev.} D}

\def\be{\begin{equation}}
\def\ee{\end{equation}}
\def\beq{\begin{equation}}
\def\eeq{\end{equation}}
\def\bea{\begin{eqnarray}}
\def\eea{\end{eqnarray}}

\def\D0{D\O }
\newcommand{\lsim}{
\mathrel{\hbox{\rlap{\hbox{\lower4pt\hbox{$\sim$}}}\hbox{$<$}}}}

\begin{document}
\vspace*{4cm}
\title{CP VIOLATION WITH $B_s$}

\author{ G. RICCIARDI}
\address{Dipartimento di Scienze Fisiche, Universit\`a di Napoli   Federico II \\ and I.N.F.N., Sezione di Napoli \\
Complesso Universitario di Monte Sant'Angelo, Via Cintia, \\ 80126 Napoli,
Italy }

\maketitle\abstracts{The observation of CP violation effects is becoming more and more significant in a variety of channels,  due to the impressive experimental effort of the last years. We  review  recent progress in $B^0_s$ semileptonic decays and in $B^0_s$ decays into CP eigenstates.}

\section{Introduction}

There are
well known   differences between the $B^0_s$ and the $B^0$ system. The
 mixing parameter $x_s \equiv \Delta m_s/\Gamma_s$ is about 30 times larger than $x_d$, and the mass and width difference are sizable.
Another important difference  is that the CP violating
 mixing phase probes  the  angle $\beta_s$  in
the unitarity triangle, which is about two order of magnitudes smaller than $\beta$ in the Standard Model,  and  hence negligibly small.
Any large variation  due to new physics can produce observable effects,
and that alone would be enough to motivate  the  study of  CP violation in
 the $B^0_s$ system.
We will review a few  decays where the observation of CP violation effects has recently  become accessible and significant,  due to the impressive experimental effort of the last years.

\subsection{Flavour-specific decays}

The  mass eigenstates can be written in terms of the flavour eigenstates
\beq
|B_{s, H}> = p \, |B^0_s> +   q \, |\bar B^0_s>  \qquad \qquad 
|B_{s, L}> = p \, |B^0_s> -   q \, | \bar B^0_s > \label{eigen111}
\eeq
where $|p|^2+|q|^2 =1$, by normalization condition.
Evidence for CP violation in $B^0_s$ mixing has been searched for,  with flavor-specific
decays, in samples where the initial flavor state is tagged.
Flavour-specific final states are
states which, due to some selection rule, can be reached directly
only by $B^0_s$ and not by $\bar{B}^0_s$ or conversely.
CP violation in the interference of mixing and decay clearly
cannot occur, as only one of the two flavour eigenstates can feed
the final state. Instead,
CP violation in the mixing and in the decay are both possible.
However, in  some  cases, the decay is dominated by a single
amplitude, and/or there are no different strong scattering phases
as required to observe CP violation in the decay.
 In that case, when the final state tag is also available, we can write the
following asymmetry
\beq  a^s_{fs}  = \frac{\Gamma (\bar B_s^0(t)
\rightarrow f) - \Gamma (B_s^0(t) \rightarrow  \bar f)}{\Gamma (\bar
B_s^0(t) \rightarrow  f)+\Gamma (B_s^0(t) \rightarrow  \bar f)} =
\frac{1-|q/p|^4}{1+|q/p|^4}
\label{asimsemi}
 \eeq
testing  the  ``wrong" final state, accessible only through mixing.
 The asymmetry $a_{fs} $  measures CP violation in mixing and
it is  independent from time  and from the final state
 (to within a sign), as it can  be ascribed to a property of the
decaying states.
In the Standard Model,  one expects
$
\left| \Gamma_{12}/M_{12} \right| \sim  m_b^2/m_t^2  \sim 10^{-3} \ll 1
$.
At lowest order in
 $|\Gamma_{12}/M_{12}|$,
 we have  \beq\left| \frac{q}{p}  \right|^2 =
1-a  \qquad \qquad  a\equiv {\rm Im} \left( \frac{\Gamma_{12}}{M_{12}} \right) =
 \frac{\Delta \Gamma_s}{\Delta m_s} \tan \phi_s \eeq
where $\phi_s \equiv  {\mathrm{ arg}} \left( -M_{12}/\Gamma_{12} \right)$, $\Delta m_s \equiv  m_{H}-m_{L} = 2 |M_{12}| $ and $
\Delta \Gamma_s =  \Gamma_{L}-\Gamma_{H}= 2 |\Gamma_{12}| \cos \phi_s$. Notice that the symbol $\phi_s$ is overloaded, since in literature it is also used  for the mixing phase induced by $M_{12}$ only.
Whatever the definition, the CP violating phase can be  related to $\beta_s$, that is  $\beta_s \equiv \textrm{arg}
\left[-{V_{tb}^\star\, V_{ts}}/{V_{cb}^\star\, V_{cs}}\right]$  in the Standard Model,  since the  dispersive
term $M_{12}$  is mainly driven by box diagrams involving virtual top quarks and
the absorptive
term $\Gamma_{12}$ is dominated by on-shell charmed intermediate
states. An additional  phase, e.~g. $\beta_s(SM) \rightarrow \beta_s(SM) + \tilde \beta_s $, it is often used to parameterize effects of new physics or non-leading hadronic contributions.

 The phase  $\phi_s \neq 0, \pi$  implies $|q/p|\neq 1$. The
  parameter $a$, that is small irrespective of the value of $\phi_s$,
implies small CP violation in the mixing.
At lowest order $a \simeq a_{fs}^s$, and
 the measured value of ${ a}_{fs}^s$ can be  translated into a constraint for both $\Delta \Gamma_s$  and $\phi_s$.
In the case of semileptonic decays, when the final state contains also a charged meson,  $a_{fs}^s$ is called semileptonic  charge asymmetry. It has been directly measured by the experiment \D0 via  the decay
$ B_s^0 \rightarrow  D_s^- \mu^+ X$~\cite{Abazov:2009wg}
\beq
a_{sl}^s= \left[ -1.7 \pm 9.1 {(\mathrm{stat})}^{+1.2}_{-2.3} ( \mathrm{syst}) \right]  \times 10^{-3}
\eeq
A related observable  is
the (like-sign) dimuon charge asymmetry ${\cal A}_{sl}^b$, which is  the difference in the number of events with a pair of positive muons minus the
number with a pair of negative muons divided by the
sum.
Since it arises from the meson mixing, if there is not a separation of the asymmetry due to $B^0$ and $B^0_s$,  ${\cal A}_{sl}^b$ can be written as
\beq
{\cal A}_{sl}^b=  C_d { a}_{sl}^d + C_s {a}_{sl}^s
\eeq
 where the coefficients depend on mean mixing probability and the production rates  of $B^0$  and $B^0_s$ mesons.
Here ${ a}_{sl}^d$ is the  semileptonic charge asymmetry in the $B^0$ system, which has been measured since 2001 at  $e^+ e^-$ machines;   the actual averaged value  is ${ a}_{sl}^d =  −0.0105 \pm  0.0064~$\cite{HFAG}.
In  2010 the experiment \D0,  with  $6\, \mathrm{fb}^{-1}$  of data,   showed evidence for anomalous ${\cal A}_{sl}^b$, deviating $3.2 \sigma$   from the SM~\cite{Abazov:2010hj}.
The 2011 \D0  update at  $9 \, \mathrm{fb}^{-1}$
shows again a  deviation, at $3.9 \, \sigma$~\cite{Abazov},  from the Standard Model value~\cite{lenz}
\beq
{\cal A}_{sl}^b = \left[ -0.787 \pm 0.172(\mathrm{stat}) \pm 0.093(\mathrm{syst}) \right] \%
\qquad \qquad 
{\cal A}_{sl}^b(\mathrm{SM}) =( -0.028^{+0.005}_{-0.006}) \%
\eeq

The extracted value for ${ a}_{sl}^s$ is in agreement with the direct determination, but improved precision or, even better,  independent measurements of semileptonic asymmetries are
needed to establish evidence of CP violation due to new
physics. The latter could come from the LHCb experiment, which has the potential for measurements of   $B^0 \rightarrow D^{\pm} \mu^{\mp} \nu$ and  $B^0_s \rightarrow D_s^{\pm} \mu^{\mp} \nu$ asymmetries.

\subsection{Decays into CP eigenstates}

The values of $\Delta \Gamma_s$
 and $\phi_s$  obtained by the semileptonic charge asymmetries have to be compared with  independent measurements  from other channels.
Particularly interesting are the so-called golden modes, which are  defined as decays where the final state is a CP eigenstate and where  all contributing Feynman diagrams carry the same CP violating phase. That ensures the absence of  CP violation in the decays, which is often plagued by large
hadronic uncertainties in the theoretical estimates. Neglecting also the small CP violation in the mixing, golden modes exhibit   interference CP violation only.
A well studied process is the  $B_s \rightarrow J/\psi \phi$ decay, whose final state
 is an admixture of different CP eigenstates, which can be disentangled through
an angular analysis of the $J/\psi (\rightarrow l^+l^-)\,  \phi (\rightarrow  K^+ K^-)$ decay products.
This decay tests directly the $B^0_s-\bar B^0_s$  mixing  phase, that  is   $\phi_M= -2  \beta_s$ in the Standard Model. In
  this channel,
the  actual world's most precise measurement of $\phi_M$ comes from the  LHCb experiment at about $ 1~\mathrm{fb^{-1}}$ of  $pp$ collisions and  it is in good agreement with Standard Model
predictions~\cite{LHCb:2011aa}.
% LHCb also provides a combination of this result with 
%an independent analysis of $B_s \rightarrow J/\psi \,  \pi \pi$ decays. 
The conflict 
 between the \D0  measurement of ${\cal A}_{sl}^b$ and the newest  LHCb
data does not appear to be theoretically  solvable with the addition of  a new  phase $\tilde \phi_M$, originated by new physics contributions to $M_{12}$,  but it seems to require non-standard additions to $\Gamma_{12}$ as well~\cite{Lenz:2012az}.

A recent player, first observed in 2011 by  the LHCb~\cite{LHCb-f0} and Belle experiments~\cite{Belle-f0},
 is  the $B^0_s \rightarrow J/\psi f_0(980)$ decay. Data have been   reported  for $B^0_s\to J/\psi f_0(980)$ with
$f_0(980)\to\pi^+\pi^-$, which is the dominant channel. LHCb has not measured the branching ratio directly, but instead its fraction, $R_
{f_0/\phi}$, with respect to the branching ratio for $B^0_s\to J/\psi \phi$ with $\phi\to K^+ K^-$.
The same ratio has been measured afterwards
by the \D0~\cite{D0} and CDF~\cite{CDF-f0}
collaborations.
All these results  are   in general agreement and point to a fraction  $R_
{f_0/\phi}$
between about 1/5 and 1/3.
The disadvantage of a
 smaller branching ratio is  compensated by the fact that the   $B^0_s\to J/\psi f_0(980)$ channel, unlike the $B^0_s\to J/\psi \phi$ one, does not require a  time-dependent angular analisys.
Indeed, because the  $f_0(980)$ is a scalar state with quantum numbers $J^{PC}=0^{++}$,  the final state of $B^0_s\to J/\psi f_0(980)$ is a $p$-wave state with the CP eigenvalue
$-1$.

 In
addition to the branching ratio
result, the CDF collaboration has reported a first measurement for the effective
$B^0_s\to J/\psi f_0(980)$ lifetime~\cite{CDF-f0}, and the LHCb collaboration has
presented a first analysis of CP violation in $B^0_s\to J/\psi f_0(980)$~\cite{LHCb-f0-CP}.
 Experimental investigations are still progressing,  leading towards more and more precise measurements of relevant observables.
It should be noted that the composition of the scalar $f_0(980)$ as a conventional $\bar q q$ meson is still under debate as of today, since alternative interpretations, e.g. as a tetraquark or a molecular state, are deemed possible.
The dominant contributions to the amplitude of $B^0_s\to J/\psi f_0$  is given by the color-suppressed tree diagram $ b \rightarrow c \bar c s$, where $f_0(980)$ is originated  by the couple $\bar s s$. Penguin and exchange diagrams give additional  contributions, that add to hadronic uncertainties. The details of the composition of $f_0(980)$ affect the amplitudes, introducing additional topologies~\cite{Fleischer:2011au}.
It becomes important to look for observables that are quite robust  with
respect to hadronic effects and thereby allow  searching  for a  large (i.e. non-standard) CP violating mixing
 phase. It has been demonstrated~\cite{Fleischer:2011au} that useful candidates in that respect are the effective lifetime of
$B^0_s\to J/\psi f_0(980)$ and the  CP violating observable $S$.
The effective lifetime  is defined as
\begin{equation}\label{lifetime-def}
 \tau_{J/\psi f_0} \equiv \frac{\int^\infty_0 t\ \langle \Gamma(B_s(t)\to J/\psi f_0(980))\rangle\ dt}
  {\int^\infty_0 \langle \Gamma(B_s(t)\to J/\psi f_0(980))\rangle\ dt}.
\end{equation}
and it  can be written in terms of $
y_s\equiv \Delta\Gamma_s/2\Gamma_s$, which in turn depends on the mixing phase.
One can investigate the dependence on the hadronic  uncertainties, finding a robust behavior under a generous range of the  parameters describing contributions from topologies different than the tree diagram \cite{Fleischer:2011au}. The dominant uncertainty   comes from the theoretical error on $\Delta \Gamma_s$ in the Standard Model.

A tagged analysis, from which we can distinguish between initially present
$B^0_s$ or $\bar B^0_s$ mesons, allows to measure the  time-dependent,
CP-violating rate asymmetry
\begin{equation}\label{t-dep-asym}
\frac{\Gamma(B_s(t)\to J/\psi f_0(980))-\Gamma(\bar B_s(t)\to J/\psi f_0(980))}{\Gamma(B_s(t)\to J/\psi f_0(980))
+\Gamma(\bar B_s(t)\to J/\psi f_0(980))}=\frac{C\cos(\Delta M_st) -
S \sin(\Delta M_st)}{\cosh(\Delta\Gamma_st/2)+
{\cal A}_{\Delta\Gamma}\sinh(\Delta\Gamma_st/2)} ,
\end{equation}
where the ``mixing-induced" CP-violating observable $S$
\begin{equation}\label{S-def}
S\equiv  \frac{- 2\,\mbox{Im}\,\lambda_{J/\psi f_0}}{1+
\bigl|\lambda_{J/\psi f_0}\bigr|^2} \qquad \lambda_{J/\psi f_0} \equiv  \frac{q}{p}\;  \frac{A(\bar B^0_s \rightarrow J/\psi f_0(980))}{A( B^0_s \rightarrow J/\psi f_0(980))}
\end{equation}
originates from interference between $B^0_s$--$\bar B^0_s$ mixing
and decay processes, and depends on the mixing phase.
The Standard Model prediction gives \cite{Fleischer:2011au}
$
\left. S(B_s^0 \to J/\psi f_0(980))\right|_{\rm SM} \in [ -0.086, -0.012],
$ and a  measurement of a sizably different $|S|$  would give us unambiguous evidence
for new physics. 
Still,  should
its value fall into the range
$
-0.1\lsim S \lsim 0,
$
the  Standard Model effects related to the hadronic parameters would preclude conclusions on
the presence or absence of CP violating new physics contributions to $B^0_{s}$
mixing.
It should be noted that the  decay  $B^0 \to J/\psi f_0(980)$, which has not yet been observed, may be used  to obtain insights into the size of such hadronic parameters. The leading contributions of the $B^0 \to J/\psi f_0(980)$ decay emerge from the $d\bar d$
component of the $f_0(980)$. Its estimated   branching ratio with $f_0(980)\to\pi^+\pi^-$ is
at the few times $10^{-6}$ level~\cite{Fleischer:2011au}, which is not  outside the reach of future experimental data taking.

New terrain 
for exploring CP violation is provided by the  $B^0_{(s)}\to J/\psi \eta^{(\prime)}$  decays. The only data come from the Belle Collaboration, that this year has given the measured values for branching fractions (of order  $\sim 10^{-4}$)  with  121.4 $\mathrm{fb}^{-1}$  of data at the  $\Upsilon(5S) $ resonance~\cite{Belle:2012aa}, following
the first observation for $B^0_{s} \to J/\psi \eta$
and the first evidence  for $B^0_{s} \to J/\psi \eta^{\prime}$  in 2009~\cite{Adachi:2009usa}. 
As before, CP violation can be investigated analyzing   the effective lifetimes and mixing-induced
CP asymmetries.
 As far as the latter are
concerned, measured values within the range $0.03 \lsim S_{J/\psi \eta^{(\prime)}}\lsim 0.09$
would not allow us to distinguish CP violating new physics contributions to $B^0_s$--$\bar B^0_s$ 
mixing from Standard Model effects, unless we can control the hadronic Standard Model corrections. 
This  can be accomplished by using e.g. the $B^0 \to J/\psi \eta^{(\prime)}$ as a control channel
and the $SU(3)_{\rm F}$ flavour symmetry. Very recently Belle has 
analyzed the branching fractions of $B^0 \to J/\psi \eta^{(\prime)}$  decays with the complete Belle data sample
of $ 772 \times 10^6$  $ B \bar B$
events collected at the $\Upsilon(4S)$ resonance  \cite{Chang:2012gn}. Only an upper limit is obtained for  $B^0\to J/\psi \eta^{\prime}$, while  the branching fractions of  $B^0\to J/\psi \eta$ is measured to be of order $O(10^{-6})$, in agreement with theoretical predictions~\cite{Fleischer:2011ib}. 
The most prominent $\eta^{(\prime)}$ decays involve photons or neutral pions in the final 
states, which is a very challenging signature for $B$-decay experiments at hadron colliders
and appears well suited for the future $e^+e^-$ SuperKEKB and SuperB projects.

\end{document}